# Optical Transfer Matrix method of layers with mechanical movements

S. Ganganagunta[a]

[a]Department of Physics, Koneru Lakshmaiah University, Guntur, Andhra Pradesh 522502,

India

Abstract

Transfer matrix method is a well-known and extensively used tool to compute the reflection and transmission coefficients of electromagnetic waves when interacting with a system of layers parallel to each other. We present here a modified form of transfer matrix method including the effects of any possible kinetic movements of layers with respect to each other with a constant velocity. We present a comprehensive analysis of the effect of velocity on the phase and amplitude of the reflection coefficient as a function of velocity. Additionally, to mimic the flow of liquids on top of layers we also present the effect of velocity gradients in the direction normal to the planar layers.

Results and Discussion

Transfer matrix method is a robust tool used in many fields such as electromagnetics, electronic circuits, nuclear reaction, acoustics etc. where a wave-like phenomenon interacts with a discrete one-dimensional objects. Specifically, in electromagnetic theory the roots of the method lies within the ray optics based Snell equations of refraction and extends to the wave nature based phenomenon such as polarization. Transfer matrix usually relates the amplitudes of scattered waves on one side of an interface to the other side. In a simple form, for a system of layers shown in Fig 1. (a) the transfer matrix at an arbitrary interface $l$ can be described as[1, 2]:

$$T^{(l)} = \alpha_l \begin{bmatrix} 1 - K_l & 1 + K_l \\ 1 + K_l & 1 - K_l \end{bmatrix} \quad (1)$$

where, $\alpha_l = 0.5$ (for Transverse Electric polarization) and $\alpha_l = 0.5 \frac{\epsilon_l}{\epsilon_{l+1}}$ (for Transverse Magnetic polarization). $K_l = k_{x;l}/k_{x;l+1}$ (for Transverse Electric polarization) and $K_l = \epsilon_{l+1} k_{x;l}/\epsilon_l k_{x;l+1}$ (for Transverse magnetic polarization). The quantity $k_{x;l}$ represent the wavevector component in the $+x$ direction in each layer given by

$$k_{x;l} = \sqrt{k_0^2 \epsilon_l - k_z^2} \quad (2)$$

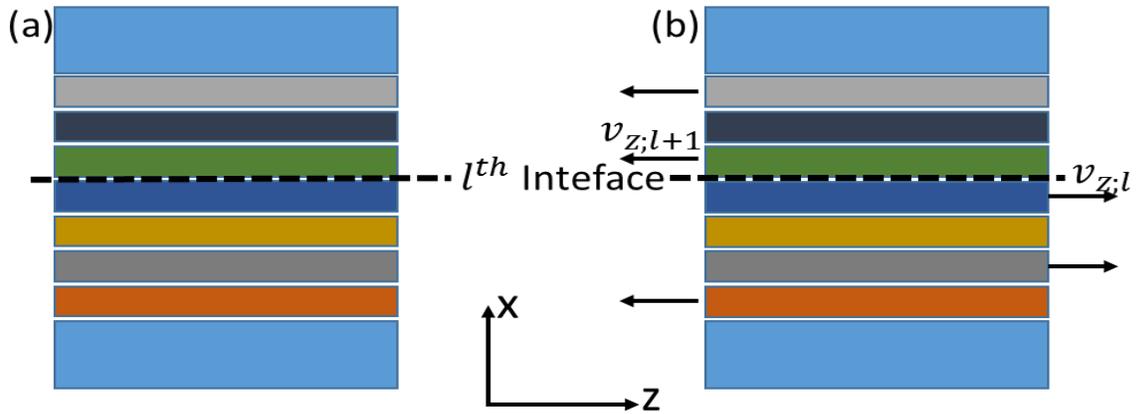

Fig. 1: (a) Schematic planar layers of a standard transfer matrix method operation. (b) Schematic geometry of mechanically moving layers to describe a modified transfer matrix.

where $k_z$ represents the incident angle as $k_z = k_0 \sin\theta$. $k_0$ is the free space wavevector.

Using modified Maxwell's equations in the presence of materials moving with uniform velocity presented in [3] and under non-relativistic approximation, the transfer matrix presented in equation (1) assuming the velocities of the $l^{th}$ and $l+1^{th}$ layers as $v_{z;l}$ and $v_{z;l+1}$ respectively as:

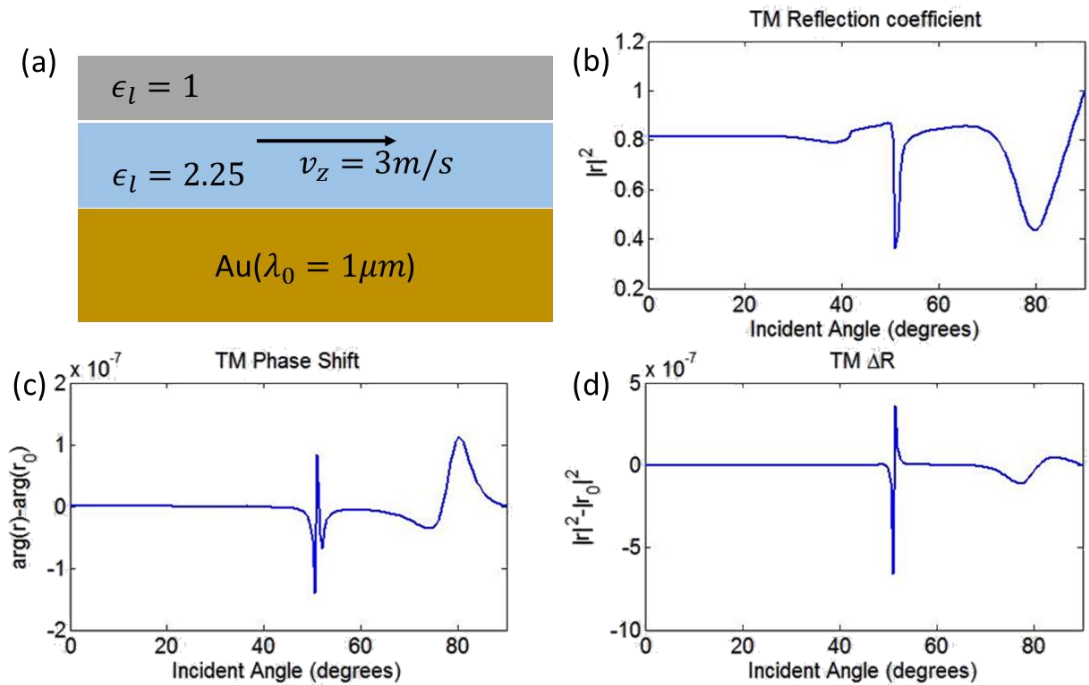

Fig. 2: (a) Schematic setup of layers. (b) Reflectance of the structure under stationary condition. (c) Variation in the phase of the reflection when the central layer moves with a constant velocity of $3\ m/s$. (d) Variation in the reflectance when the central layer moves with a constant velocity of $3\ m/s$.

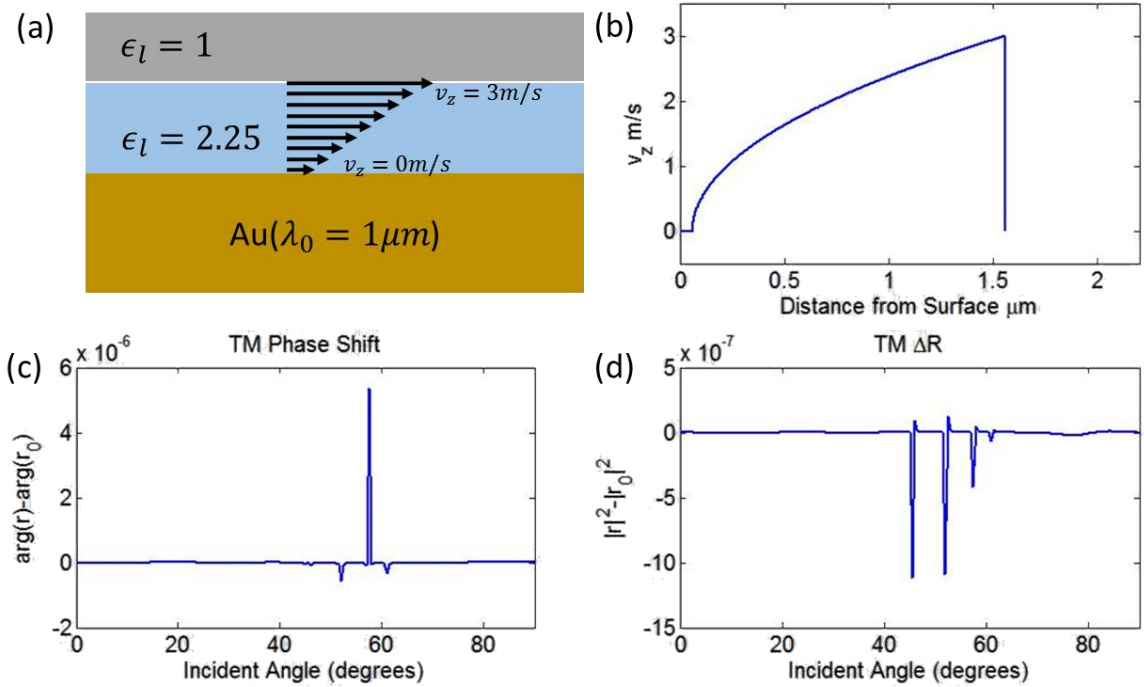

Fig. 3 (a) Schematic setup of layers with a gradient velocity profile. (b) Velocity profile from bottom to top of the layer. (c) Variation in the phase of the reflection with the velocity gradient. (d) Variation in the reflectance with velocity gradient.

$$T_v^{(l)} = \alpha_l \begin{bmatrix} 1 - K'_l & 1 + K'_l \\ 1 + K'_l & 1 - K'_l \end{bmatrix} \quad (3)$$

where $K'_l$ has identical definitions as before for TE and TM polarizations. However, the dispersion in equation (2) now modifies as:

$$k_{x;l} = \sqrt{k_0^2 \epsilon_l - \left(k_z + k_0 \beta_{z;l}(\epsilon_l - 1)\right)^2} \quad (4)$$

where $\beta_{z;l}$ represents the relative velocity of the layer as $\beta_{z;l} = v_{z;l}/c$.

Fig. 2 demonstrates the effect of motion of a single dielectric layer on top of a metal substrate. It can be clearly observed that under constant motion of the layer a very small but reasonable shift happens in both phase and magnitude of the reflection coefficient of the system. However, such a mechanical motion is unphysical and involves a high degree of static and kinetic frictions between the layers. The developed transfer matrix can be applied in a more realistic scenario where a layer of liquid moves on top of a metal substrate. In contrast to solid layers, a liquid layer will have viscous effects that restricts the velocity of the liquid exactly at interface to be zero[4]. This restriction creates a phase gradient in the velocity profile. To perform the computation we divide the central layer in a large number of layers assuming each propagating with a constant velocity but represent a gradient overall. The results are summarized in Fig. 3. Clearly, the phase response of the reelection coefficient under velocity gradient is different from constant velocity. In addition, the magnitude of the phase difference is a magnitude higher in order compared to constant velocity.

In conclusion, we developed a modified transfer matrix approach to compute electromagnetic response of planar layered materials that are under mechanical motion with constant velocity. The calculations from the method demonstrate minor variations in the phase and amplitude of the reelection coefficients of layers under motion. The phase variation is observed to be relatively higher when gradient of the velocity considered in the place of a constant profile. Such cross-polarized reflection measurements are very useful in bio-medical imaging techniques [5-11]. The developed formulation could trigger novel opto-mechanical devices.